\documentclass[epj,nopacs]{svjour}
\usepackage[T1]{fontenc}
\usepackage[latin9]{inputenc}
\usepackage{color}
\usepackage{array}
\usepackage{amstext}
\usepackage{graphicx}
\usepackage{esint}
\usepackage{graphics}
\begin{document}
\title{Gluon fragmentation into S-wave heavy quarkonium}
\author{S. Mohammad Moosavi Nejad\inst{1,2}, Delpasand Mahdi\inst{1}}

\institute{Faculty of Physics, Yazd University, P.O. Box
89195-741, Yazd, Iran \and
School of Particles and Accelerators,
Institute for Research in Fundamental\\
Sciences (IPM), P.O.Box
19395-5531, Tehran, Iran}

\date{Received: date / Revised version: date}
%
\abstract{Fragmentation is the dominant production mechanism for heavy  hadronic bound 
states  with large transverse momentum. We numerically calculate the initial
$g\rightarrow H(Q\bar{Q})$ fragmentation functions (FFs) using the nonrelativistic QCD
factorization approach. Our analytical expression of FFs depends on both the momentum fraction $z$
and the transverse momentum of the gluon, and 
contains  most of the kinematical and dynamical properties of the process.
Specifically, using the perturbative QCD we present the FF for a gluon 
to split into $S$-wave charmonium meson $H_c$ to leading order in the QCD coupling constant.}

\maketitle

\section{Introduction}
\label{sec:intro}

Heavy quarkonia, as the bound states of a heavy quark and antiquark are the simplest
particles when the strong interactions are concerned. 
Heavy quarkonium production in hard scattering collisions has a long history of 
theoretical calculations and experimental measurements \cite{Brambilla:2010cs} and intense efforts
are expected to continue as  the Large Hadron Collider (LHC) makes  data available
with unprecedented momentum transfers.\\
In production of heavy quarkonia with sufficiently large transverse
momentum $k_T$, the dominant mechanism is actually \textit{fragmentation}
while the direct leading order production scheme is normally suppressed. 
Fragmentation refers to the process of a 
parton with high transverse momentum which subsequently decays to form a jet containing 
the expected hadron \cite{Braaten:1994xb}. The $Q\bar{Q}$ pair is created with a separation of
order $1/m_Q$, where $m_Q$ stands for the mass of the heavy quark $Q(=c, b)$.
The lowest states in the charmonium ($\eta_c, J/\psi$) and bottomonium ($\Upsilon$)  systems have typical radius that are 
significantly smaller than those of hadrons containing light quarks.
They have simple internal structures, consisting primarily of a nonrelativistic 
quark and antiquark, so in recent years a great deal of theoretical effort has been focused
on the nonrelativistic QCD factorization approach \cite{Bodwin:1994jh} to calculate the quarkonium 
production rates.

Generally, according to the factorization theorem of QCD \cite{collins} 
the production of heavy quarkonium $H$ in the typical scattering process of $A+B\rightarrow H(k_T)+X$, can be expressed as
\begin{eqnarray}
d\sigma&=&\sum_{a,b,c}\int_0^1 dx_a\int_0^1 dx_b\int_0^1 dz f_{a/A}(x_a, \mu)f_{b/B}(x_b, \mu)\times\nonumber\\
&&d\hat\sigma(a+b\rightarrow c+X)D_{c\rightarrow H}(z, \mu),
\end{eqnarray}
where $\mu$ is a factorization scale, $a$ and $b$ are incident partons in the colliding initial hadrons $A$ and $B$ respectively,
$f_{a/A}$ and $f_{b/B}$ are the parton distribution functions at the scale $\mu$, $c$ is the
fragmenting parton (either a gluon or a quark) and $X$ stands for the unobserved jets. 
Here, $D_{c\rightarrow H}(z, \mu)$ is the fragmentation function(FF) at the scale $\mu$ which
can be obtained by evolving from the initial FF $D_{c\rightarrow H}(z, \mu_0)$
using the Dokshitzer-Gribov-Lipatov-Altarelli-Parisi (DGLAP) renormalization group equations \cite{dglap}
\begin{eqnarray}\label{dglap}
\frac{d}{d\ln\mu^2}D_{i\rightarrow H}(z, \mu)=\sum_j\int_z^1\frac{dy}{y}P_{ij}(\frac{z}{y}, \mu)D_{j\rightarrow H}(y, \mu),\nonumber\\
\end{eqnarray} 
where $P_{ij}$ are the Altarelli-Parisi splitting functions  for the splitting of the parton of the type $i$
into a parton of the type $j$ with longitudinal momentum fraction $x$. 
The initial scale FF $D_{i\rightarrow H}(z, \mu_0)$ is a universal function which can be 
obtained phenomenologically or analytically. 
The universality of the initial scale FFs, first was suggested  in \cite{Mele}
in the framework of  $e^-e^+$ annihilation and afterward  was proved in a more
general way in Ref.~\cite{Cacciari:2001cw}.
The importance of FFs is for the model independent predictions of the cross
sections at the LHC in which a hadron is detected in the outgoing
productions as a colorless bound state.\\
In \cite{Nejad:2013vsa}, using the perturbative QCD we calculated the initial scale FF for c-quark to split into
S-wave $D$-meson considering  the effect of hadron mass. There, we compared our result with the current well-known phenomenological
models and we also compared the FF with experimental data from BELLE and CLEO. Our result was in good consistency with the other ones.\\
In Ref. \cite{MoosaviNejad:2011yp} we studied the importance of gluon FF by accounting the effect of gluon fragmentation on
the scaled-energy distribution ($x_B$) of bottom-flavored hadrons B inclusively produced in top-quark
decays  in the standard model ($t\rightarrow bW^+(+g)\rightarrow BW^++X$)
and in the general two Higgs doublet model ($t\rightarrow bH^+(+g)\rightarrow BH^++X$). We found that 
gluon fragmentation leads to an appreciable reduction in the partial decay
width at low values of $x_B$ and for higher values of $x_B$ the NLO result is practically exhausted by
the $b\rightarrow B$ contribution.\\
In this paper we calculate the gluon FF into the S-wave heavy quarkonium H ($D_g^{H}(z, \mu_0)$)
by calculating a specific physical process in perturbative QCD
in the finite momentum frame of the fragmenting gluon. Specifically, we focus on  
the S-wave charmonium $H_c$ (i.e. $\eta_c, J/\psi$) as a heavy quarkonium and
 present our result for the $D_g^{H_c}(z, \mu_0)$-FF, where the $z$ is defined in the Lorentz boost
invariant forms as \cite{Qi:2007sf}
\begin{eqnarray}
z&=&\frac{E^H+k_L^H}{E^g+k_L^g},\\
z&=&\frac{k^g.k^H}{(k^g)^2},\\
z&=&\frac{\sqrt{M_H^2+(k_L^H)^2}}{E^g},\\
z&=&\frac{\sqrt{M_H^2+(k_L^H)^2}+k_L^H}{E^g+k_L^g},\\
z&=&\frac{k_L^H}{k_L^g}.
\end{eqnarray}
In the above equations we take the z-axis along the momentum of outgoing meson and 
$E^g, k_L^g, E^H$ and $k_L^H$ are the energies and longitudinal components (z-components) of the four-momenta of
the fragmenting gluon and the produced heavy quarkonium $H$, respectively. The first definition (3) is the usual light-cone form.
The first and second definitions of $z$ are hard to be employed in the application of the gluon FFs,
 because they involve the longitudinal momentum of the resulting heavy quarkonium.
Instead, usually the non-covariant definitions (Eqs. (5)-(7)) are used approximately, which are convenient for
the finite momentum frame. In \cite{Qi:2007sf}, authors analyzed the uncertainties induced by different
definitions of the $z$ in the application of gluon to heavy quarkonium FF. They calculated the initial $g\rightarrow J/\psi$ FF by
calculating the differential cross section of process $g+q\rightarrow q+g^\star(\rightarrow J/\psi+g+g)$.
They showed that the FFs have strong dependence on the gluon momentum $\vec{k}$, and
when $|\vec{k}|\rightarrow \infty$ these FFs approach to the FF in the light-cone definition (3) 
and large uncertainties remain while the non-covariant definitions of $z$ are employed.

The FFs are related to the low energy part  of the hadron production
and they consist of the nonperturbative aspects of QCD.   
There are two main approaches for evaluating the initial scale FFs.
In the first approach called the phenomenological approach,
 these functions are extracted from experimental data analysis instead of theoretical calculations.
This scheme is explained in more detail in section \ref{sec:one}.
The second approach is based on the fact that  the FFs for mesons containing, at least, 
a heavy quark can be calculated theoretically using
perturbative QCD (pQCD) \cite{Ma:1997yq,Braaten:1993rw,Chang:1991bp,Braaten:1993mp,Scott:1978nz}.
We employ this approach to drive an exact analytical
form of FF for the transition of $g\rightarrow H_c$.

This paper is organized as follows.
In Sec.~\ref{sec:one}, we explain the phenomenological approach to calculate the FFs and  
introduce some well-known phenomenological models.
In Sec.~\ref{sec:two}, the theoretical approach  to calculate the FFs is introduced in detail.
We discuss the use of pQCD in calculating  
 the fragmentation of a charm quark into the heavy charmonium $H_c$ 
and in Sec.~\ref{sec:four},   our conclusion is summarized.

\section{Phenomenological determination of FFs}
\label{sec:one}

The FFs describe hadron production probabilities from the initial partons and
their importance is for model independent predictions of the cross
sections and decay rates at the LHC. They can also be applied to detect the internal structure of the
exotic hadrons using the differences between the disfavored and favored FFs \cite{Hirai:2010cs}.
One of the main approaches to evaluate the FFs, which is normally called the phenomenological approach,
is based on the experimental data analyzing.
In this approach, the FFs are mainly determined by hadron production data analysis
of $e^-e^+$ annihilation, lepton-hadron deep inelastic
scattering (DIS) and hadron-hadron scattering 
processes by working either  in Mellin-N space  \cite{Nason:1999zj,Cacciari:2005uk}
or in $x$-space \cite{Albino:2005me,Kniehl:2000fe,Kneesch:2007ey}.
Among these methods, the FFs are mainly  determined by hadron production data of $e^-e^+$ annihilation, because
there are more accurate data for this process.\\
In this approach, according to  Collin's factorization theorem \cite{collins}
 the cross section of hadron production in the $e^-e^+$ annihilation is expressed by the convolution
of  partonic hard-scattering cross sections ($e^-e^+\rightarrow q\bar q(+g)$), 
 and  the nonperturbative FFs $D_i^H(z, Q^2)$, describing the transition
of a parton into an outgoing hadron $H$,
\begin{eqnarray}\label{fac}
\frac{d}{dz}\sigma(e^+e^-\rightarrow HX)=\sum_{i=g, u, d, s,\cdots} C_{i}(z, \alpha_s) \otimes D_i^H(z, Q^2),\nonumber\\
\end{eqnarray}
where, $C_i(z, \alpha_s) $ are the Wilson coefficient functions based on the partonic cross sections 
which are  calculated in the perturbative QCD \cite{Kniehl:2000fe,Nason:1993xx}, and the convolution
integral is defined as $f(x)\otimes g(x)=\int_x^1 dy/y f(y)g(x/y)$. Here, $z=2E_H/\sqrt{s}$ is the  
 fragmentation parameter where  $E_H$ is the  energy of observed hadron and $s=Q^2$ is the squared center-of-mass energy.
In fact, the fragmentation parameter  $z$ refers  to the energy fraction of process
which is taken away by the detected hadron.\\
In the phenomenological scheme, the FFs are parameterized in a convenient functional form at  the initial
scale $\mu_0$ in each order, i.e. leading order (LO) and next-to-leading order (NLO). 
The initial scale $\mu_0$ is different for partons and the initial FFs
 are evolved to the experimental $\mu$ points by the DGLAP equations (\ref{dglap}).
The FFs are parameterized in terms of a number of free parameters which are fixed by an $\chi^2$ analysis of the
$e^+e^-\rightarrow H+X$ data at the scale $\mu^2=s$. In \cite{Soleymaninia}, using this scheme  
we presented a new functional form of $\pi^+/K^+$ FFs up to next-to-leading order
through a global fit to single-inclusive electron-positron annihilation data.
The situation is very similar to the one for determination of PDFs.

\begin{figure}
\begin{center}
\includegraphics[width=1\linewidth,bb=37 192 552 629]{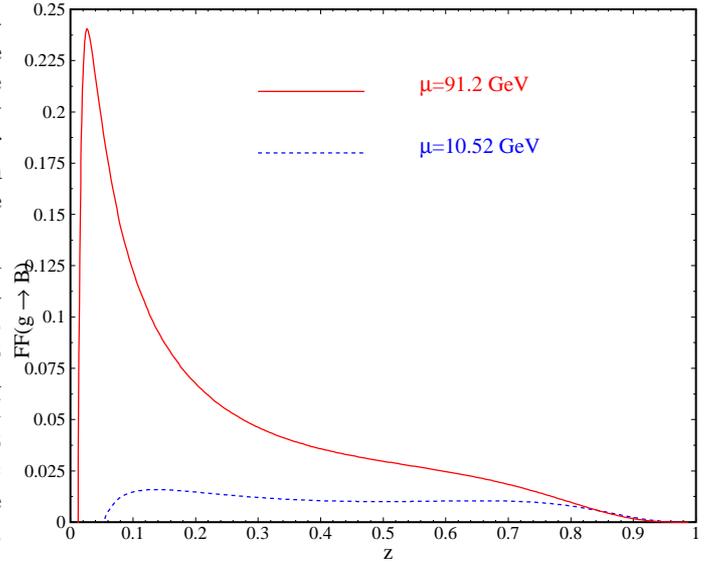}
\caption{\label{bmeson}%
$g\rightarrow B$ FF at the scales $\mu=10.52$~GeV and $\mu=m_Z$ using the power model.}
\end{center}
\end{figure}

\begin{figure}
\begin{center}
\includegraphics[width=1\linewidth,bb=37 192 552 629]{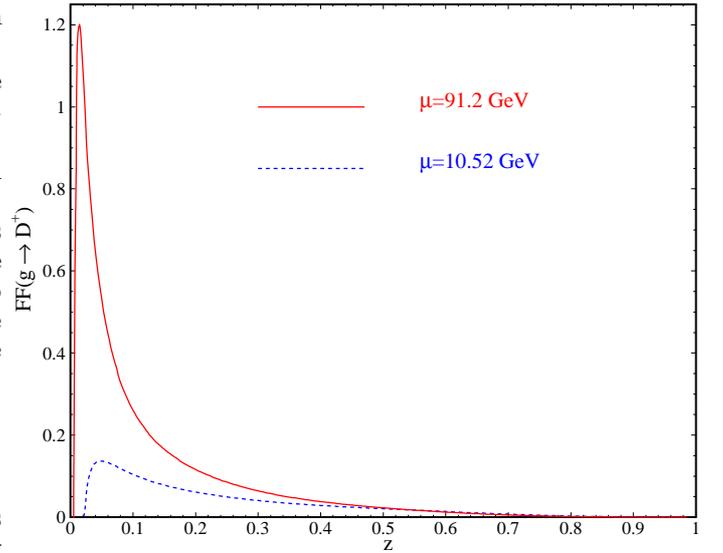}
\caption{\label{dplusmeson}%
$g\rightarrow D^+$ FF at the scales $\mu=10.52$~GeV and $\mu=m_Z$ using the Bowler model.}
\end{center}
\end{figure}

Various phenomenological models like Peterson model \cite{Peterson:1982ak}, Lund model \cite{Andersson:1983ia},
Cascade model \cite{Webber:1983if} etc., have been developed to describe the fragmentation processes.
In \cite{Kniehl:2011bk}, authors reported  the 
nonperturbative $B$-hadron  FFs that were determined at NLO in
the ZM-VFN scheme through a joint fit to
$e^+e^-$-annihilation data taken by ALEPH \cite{Heister:2001jg} and OPAL
\cite{Abbiendi:2002vt} at CERN LEP1 and by SLD \cite{Abe:1999ki} at SLAC SLC.
Specifically, the power ans\"{a}tze $D(z,\mu_F^\text{ini})=Nz^\alpha(1-z)^\beta$
with three free parameters 
was used as the initial condition for the $b\to B$ FF at
$\mu_F^\text{ini}=4.5$~GeV, while the gluon and light-quark FFs were generated
via the DGLAP evolution. The fit yielded $N=4684.1$, $\alpha=16.87$, and $\beta=2.628$.
In Ref.~\cite{Kneesch:2007ey}, authors
determined the FFs for $D^0, D^+$ and $D^{\star +}$ mesons by fitting the data from
the BELLE, CLEO, ALEPH, and OPAL Collaborations in the modified minimal-subtraction ($\overline{MS}$) factorization scheme  by
considering the model suggested by Bowler \cite{Bowler}, as $D_q^{H_c}(z,\mu_0)=Nz^{-(1+\gamma^2)}(1-z)^a e^{-\gamma^2/z}$.
At the scale $\mu=m_c=1.5$~GeV, the Bowler model is taken for the $c$-quark FF, while the FFs of the light
quarks q ($q=u, d, s$) and the gluon are set to zero. Then these FFs are evolved to higher
scales using the DGLAP equations at NLO.\\
In Figs. \ref{bmeson} and \ref{dplusmeson}, the behavior of $g\rightarrow H(=B, D^+)$ FFs at the scales $\mu=10.52$~GeV and $\mu=m_z=91.2$~GeV
are shown. The scale of $\mu=10.52$~GeV, which is much close to the production threshold of D-mesons, has been set as the center-of-mass
energy of $e^-e^+$ annihilation by the Belle and the CLEO Collaborations \cite{Seuster:2005tr}. 

\section{Theoretical determination of FFs}
\label{sec:two}

\begin{figure}
\begin{center}
\includegraphics[width=0.5\linewidth,bb=199 510 392 729]{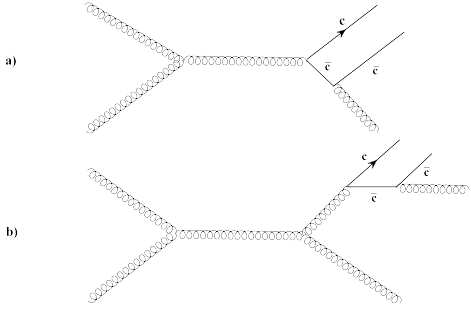}
\caption{\label{feynman}%
Feynman diagrams that contributes to charmonium production: (a) $gg\rightarrow c\bar{c}g$ at order-$\alpha_s^3$, 
(b) $gg\rightarrow c\bar{c}gg$ at order-$\alpha_s^4$.}
\end{center}
\end{figure}

\begin{figure}
\begin{center}
\includegraphics[width=0.4\linewidth,bb=199 452 392 729]{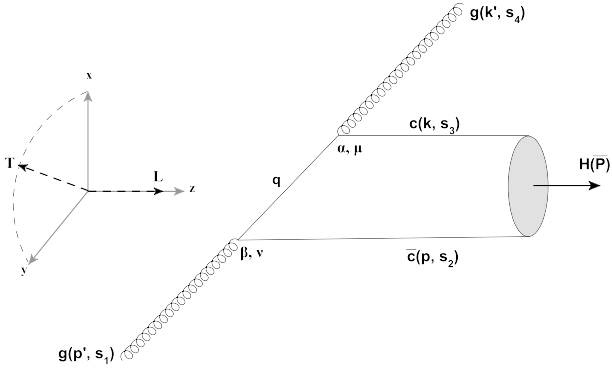}
\caption{\label{ff}%
Formation of a heavy quarkonium. A gluon forms a bound state $c\bar c$
with a gluon produced through a single c-quark.}
\end{center}
\end{figure}

The second current approach for calculating the FFs is based on the fact that 
the FFs for hadrons containing a heavy quark can be computed theoretically using
perturbative QCD (pQCD) \cite{Ma:1997yq,Braaten:1993rw,Chang:1991bp,Braaten:1993mp,Scott:1978nz}.
The first theoretical attempt to explain the production procedure of hadrons containing a heavy quark
or a heavy antiquark  was made by Bjorken \cite{Bjorken:1977md} by using a
naive quark-parton model (QPM). He construed that the inclusive distribution of  heavy hadron  should peak
nearly at $z=1$, where $z$ stands for  the scaled energy variable. This property
is mainly important for heavy quarks for which the peak of heavy quark FF occurs closer to $z=1$.
In following, Peterson  \cite{Peterson:1982ak} proposed the popular form of FF
which manifestly behaves as $(1-z)^2$ at large $z$ values, using a nonrelativistic quantum mechanical parton model.
The perturbative QCD approach  was followed by Suzuki \cite{Suzuki:1977km},  Ji and Amiri \cite{Amiri:1986zv}. While
in this approach Suzuki calculates the heavy FFs using a Feynman diagram similar to that in Fig.~\ref{ff}.\\
Here, we focus on gluon fragmentation into a heavy quarkonium  considering a special 
example: $g\rightarrow H_c(=\eta_c, J/\psi)$, and drive an exact analytical
form of  $D_g^{H_c}(z, \mu_0)$ using the Suzuki's approach which embeds most of the kinematical and
dynamical properties of the process. Our result can be directly used for the bottomonium state ($\Upsilon$-system)
with some simple replacements.\\
In a hadron collider, a charmonium meson with large transverse momentum $k_T$ can either be produced
directly at large $k_T$ or  can be produced indirectly by the decay of a $B$ meson or a higher charmonium state with
large $k_T$. A typical Feynman diagram which contributes to the production of the
charmonium state at the order-$\alpha_s^3$ is shown in Fig. \ref{feynman}a and the order-$\alpha_s^4$ radiative corrections to this 
process is shown in Fig. \ref{feynman}b. In most regions of phase space, the 
virtual gluons in Fig. \ref{feynman}b are off their mass shells by values of order $k_T$, and
the contribution from this diagram is suppressed relative to the diagram in Fig. \ref{feynman}a by
a power of the strong coupling constant $\alpha_s(k_T)$.

Using the Suzuki's  approach which is based on the perturbative QCD scheme, we obtain the analytical form of 
FF for gluon to split into the charmonium state   
considering the Feynman diagram for $g\rightarrow H(c\bar c)+g$ in the order of $\alpha_s^2$.
This diagram along with the spins and the four-momenta of meson and partons is shown in Fig.~\ref{ff}.

Following Ref.~\cite{Suzuki:1985up}, we adopt the infinite momentum frame where
the fragmentation  parameter in the usual light-cone form (3) is reduced to the more popular form $z=E^H/E^g$ (5).
We also  neglect the relative motion of  $Q$ and $\bar Q$ therefore we assume, for simplicity, that
$Q$ and $\bar Q$ are emitted collinearly with each other and they  move  along the $z$-axes.
Therefore we set the relevant four-momenta in Fig.~\ref{ff} as
\begin{eqnarray}\label{kinematic}
p_\mu^\prime =[p_0^\prime, k_T, p_L^\prime] &&\quad      p_\mu=[p_0, \vec{0}, p_L] \nonumber\\
k_\mu^\prime =[k_0^\prime, k_T, k_L^\prime]  &&\quad     k_\mu=[k_0, \vec{0}, k_L]\nonumber\\
&&\hspace{-1cm}\bar P_\mu=[\bar P_0, \vec{0}, \bar P_L],
\end{eqnarray}
where for the  momentum of the produced meson one has  $ \bar{P}_L=p_L+k_L$.
Considering the definition of fragmentation  parameter, $z=E^H/E^g=\bar P_0/p_0^\prime$, we
also may write the parton energies in terms of the  initial gluon energy $p_0^\prime$  as:
$p_0=x_1 z p_0^\prime, k_0=x_2 z p_0^\prime,  k_0^\prime=(1-z) p_0^\prime$, 
where $x_1=p_0/\bar{P}_0$ and $x_2=k_0/\bar{P}_0$ are  the meson energy fractions carried by the constituent quarks.
Following Ref.~\cite{GomshiNobary:1994eq}, we also assume  that the contribution of each constituent quark from the meson energy is proportional
to its mass, i.e. $x_1=m_c/M$ and $x_2=m_{\bar c}/M$ where $M=m_c+m_{\bar c}=2m_c$.\\
Considering the four-momenta in Fig. \ref{ff}, we start with the definition of FF introduced in Refs.~\cite{Suzuki:1985up,GomshiNobary:1994eq} as
\begin{eqnarray}\label{first}
D_{g\rightarrow H_c}(z, \mu)=\int d^3\vec{p} d^3\vec{k} d^3\vec{k^\prime} \overline{|T_M|^2} \delta^3(\vec{k^\prime}+\vec{p}+
\vec{k}-\vec{p^\prime}),\nonumber\\
\end{eqnarray}
where the average probability amplitude squared $\overline{|T_M|^2}$ is calculated 
as $\sum_{s}|T_M|^2 /(1+2s_g)$ in which  the summation is going
over the spins and colors  and   $s_g$ is the initial gluon spin. 
The probability amplitude to split a gluon into the meson ($T_M$)
is expressed as the convolution of the hard scattering 
amplitude $T_H$ which is, in essence, the partonic cross section to produce a heavy quark-antiquark ($Q\bar Q$)
pair with certain quantum numbers, and the process-independent distribution amplitude $\Phi_M$, i.e. 
\begin{eqnarray}\label{base}
T_M=\int [dx_i] T_H(x_i,Q^2) \Phi_M(x_i, Q^2),
\end{eqnarray}
where $[dx_i]=dx_1dx_2\delta(1-x_1-x_2)$.
The short-distance coefficient $T_H$ can be calculated as perturbation series in the strong coupling constant $\alpha_s$.
The long-distance distribution amplitude $\Phi_M$ which contains the bound state nonperturbative dynamic of
produced meson, is the probability amplitude for a $Q\bar Q$ pair  to evolve
into a particular heavy quarkonium state.
 The probability amplitude $\Phi_M$ is related to the mesonic wave function $\Psi_M$ by
 \begin{eqnarray}\label{formul}
\Phi_M(x_i, Q^2)&=&\int 2 (2\pi)^3\delta\big[\sum_{j=1}^2  \vec{q_{\bot j}}\big]\prod_{i=1}^2\frac{d^2 \vec{q_{\bot i}}}{2(2\pi)^3}\times\nonumber\\
&&\Psi_M(x_i, \vec{q_{\bot i}})\Theta( \vec{q_{\bot i}}^2<Q^2),
\end{eqnarray}
where, $\Theta(x)$ is the Heaviside step function 
and  $ \vec{q_{\bot i}}$ stands for  the  transverse momentum of constituent quarks. 
A simple nonrelativistic wave function is given as \cite{brodsky}
\begin{eqnarray}\label{formula}
\Psi_M(x_i, \vec{q_{\bot i}})=\frac{(128 \pi^3 b^5 M)^{\frac{1}{2}}}{x_1^2 x_2^2\big[M^2-\frac{m_1^2+\vec{q_{\bot 1}}^2}{x_1}-
\frac{m_2^2+\vec{q_{\bot 2}}^2}{x_2}\big]^2},\nonumber\\
\end{eqnarray} 
 where $M$ is the meson mass and $b$ is the binding energy of the mesonic bound state.
Working in the infinite-momentum frame  we integrate 
over $\vec{q_{\bot i}} (0\leq\vec{q_{\bot i}}^2\leq \infty)$ where $\vec{q_{\bot i}}$ stands for either $\vec{q_{\bot 1}}$ or $\vec{q_{\bot 2}}$.
 The integration yields
 \begin{eqnarray}
\Phi_M(x_i, Q^2)=\frac{(128 \pi b^5 M)^{\frac{1}{2}}}{16\pi^2(x_1+x_2) (m_1^2 x_2+m_2^2 x_1-x_1 x_2 M^2)},\nonumber\\
\end{eqnarray} 
 which grows rapidly  at $x_1=1-x_2=m_1/M$ when $M$ is set to $m_1+m_2$. Therefore this function can be
 estimated as a delta function \cite{Amiri:1985mm}. In conclusion, 
the probability amplitude  for a S-wave pseudoscalar meson at large $Q^2$, reads
\begin{eqnarray}
\Phi_M\approx\frac{f_M}{2\sqrt{3}} \delta(x_1-\frac{m_1}{m_1+m_2}),
\end{eqnarray}
where $f_M$ refers to the decay constant for the meson.
The delta-function form is convenient for our assumption where we ignore the relative motion of quark and antiquark and
thus the constituent quarks are emitted collinearly with each other and they have no transverse momentum.
However, as in Ref.\cite{Bodwin:2012xc}  mentioned, the squared relative velocity of the heavy
quark and the heavy antiquark in the quarkonium rest frame is $v^2\approx 0.22$ for the $J/\psi$ and $v^2\approx 0.1$ for the $\Upsilon$.
These theoretical results are generally in consistency with experimental measurements of quarkonium production cross sections.

Considering Fig. \ref{ff}, in which we make a leading-order approximation to form the $J/\psi$-meson,
the squared  QCD amplitude $|T_H|^2$ is expressed as
\begin{eqnarray}\label{forth}
\overline{|T_H|^2}=\frac{m_c^2m_{\bar c}^2}{8p_0 k_0 k_0^\prime p_0^\prime}\frac{\overline{|\textbf{\textit{M}}|^2}}{(p_0+k_0+k_0^\prime-p_0^\prime)^2},
\end{eqnarray}
where, using the feynman rules the transition amplitude $\textbf{\textit{M}}$ reads
\begin{eqnarray}
\textbf{\textit{M}}=\frac{g_s^2 C_F}{(k+k^\prime)^2-m_c^2}\{\bar{u}(k, s_3)\Gamma v(p, s_2)\}.
\end{eqnarray}
Here, $\Gamma=\displaystyle{\not}\epsilon_4^\star(\displaystyle{\not}k+\displaystyle{\not}k^\prime+m_c)\displaystyle{\not}\epsilon_1$
where $\epsilon$ is the polarization vector of gluon, $C_F$ 
is the color factor and $g_s$ is  the strong coupling constant. Performing
an average over the initial spin states and a sum over the final spin states, the mean amplitude squared reads
\begin{eqnarray}
\overline{|\textbf{\textit{M}}|^2}&=&\frac{1}{1+2s_1}\sum_{s}\textbf{\textit{M}} \textbf{\textit{M}}^\star\nonumber\\
&&=\frac{g_s^4 C_F^2}{[(k+k^\prime)^2-m_c^2]^2}\sum_s L^{\mu\nu}L_{\mu\nu},
\end{eqnarray}
where, using $\sum_s \epsilon^{\mu\star}(p, s)\epsilon^\nu(p, s)=-g^{\mu\nu}$ we have
\begin{eqnarray}\label{tensor}
\tilde{\omega}&=&\sum_s L^{\mu\nu}L_{\mu\nu}=\sum_{s_1, s_4}Tr\{\Gamma(\displaystyle{\not}p-m_{\bar c})\bar\Gamma(\displaystyle{\not}k+m_c)\}\nonumber\\
&&=32(p.k^\prime)(k.k^\prime)-32m_c^2(p.k)-32m_c^2(p.k^\prime)\nonumber\\
&&-64m_c^2(k.k^\prime)-64m_c^4.
\end{eqnarray}
To obtain the FF for an unpolarized meson, considering  (\ref{first}-\ref{tensor}) we have
\begin{eqnarray}
D_{g\rightarrow H_c}(z, \mu_0)&=&(\frac{m_c^4g_s^4C_F^2}{24})(\frac{f_M^2}{12})
\int\frac{\tilde{\omega}d^3\vec{ k^\prime}}{[(k+k^\prime)^2-m_c^2]^2}
\nonumber\\
&&\times \int\frac{d^3\vec{k}}{k_0 p_0^\prime k_0^\prime}\int \frac{d^3\vec{p}\delta^3(\vec{k}+\vec{p}+
\vec{k^\prime}-\vec{p^\prime})}{D_0^2p_0},\nonumber\\
\end{eqnarray}
where $D_0=p_0+k_0+k_0^\prime-p_0^\prime$ is the energy denominator.\\
To perform the phase space integrations we consider the following integral
\begin{eqnarray}
&&\int \frac{d^3\vec{p}\delta^3(\vec{k}+\vec{p}+
\vec{k^\prime}-\vec{p^\prime})}{p_0 D_0^2}\nonumber\\
&&=\frac{p_0}{[m_c^2+(k+k^\prime)^2+2p.(k+k^\prime)]^2},
\end{eqnarray}
where, considering (\ref{kinematic}) one has
\begin{eqnarray}
(k+k^\prime)^2&=&m_cM[\frac{m_c}{M}+\frac{k_T^{2}}{M^2}\frac{z}{1-z}+\frac{1}{z}-1],\nonumber\\
p.k&=&m_c^2,\nonumber\\
p.k^\prime &=&\frac{m_czk_T^2}{2M(1-z)}+\frac{m_cM(1-z)}{2z}.
\end{eqnarray}
We also write 
\begin{eqnarray}
\int{d^3\vec{k^\prime}f(z, \vec{k_T^{2}})}\approx k_0^\prime f(z, \left\langle k_T^{2}\right\rangle),
\end{eqnarray}
where, for simplicity, we replaced the transverse momentum integration by its average value $ \left\langle k_T^2 \right\rangle$, 
which is a free parameter and can be specified experimentally.\\
Putting all in (\ref{first}) we obtain the fragmentation function as
\begin{eqnarray}\label{last}
D_{g\rightarrow H_c}(z, \mu_0)&=&\frac{N z}{F(z, \left\langle k_T^{2}\right\rangle)}\times\nonumber\\
&&\hspace{-1cm}\bigg\{\bigg[\frac{z^2k_T^{2}+M^2(1-z)^2}{Mz(1-z)}\bigg]^2(1-6m_c)-12m_c^2\bigg\},\nonumber\\
\end{eqnarray}
where,
\begin{eqnarray}
F(z, \left\langle k_T^{2}\right\rangle)&=&\bigg[2\frac{m_c}{M}+\frac{k_T^{2}}{M^2}\frac{z}{1-z}+\frac{1}{z}-1\bigg]^2\nonumber\\
&&\times \bigg[\frac{k_T^{2}}{M^2}\frac{z}{1-z}+\frac{1}{z}-1\bigg]^2,
\end{eqnarray}
and $N$ is proportional to $(\pi C_F \alpha_s f_M)^2$ but it is related to the normalization condition
$\int_0^1 D_{g\rightarrow H_c}(z, \mu_0) dz=1$  \cite{Amiri:1986zv,Suzuki:1985up}.

\begin{figure}
\begin{center}
\includegraphics[width=1.0\linewidth,bb=88 620 322 769]{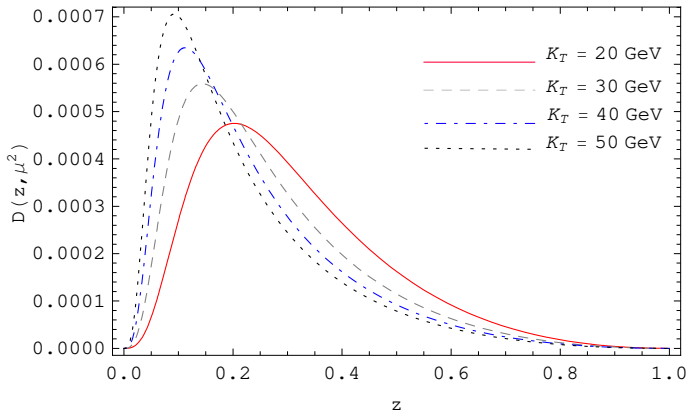}
\caption{\label{plot}%
$g\rightarrow H_c$ FF as a function of $z$ and $k_T$ at the scale $\mu=2 m_c$.}
\end{center}
\end{figure}

In general, fragmentation function $D_{g\rightarrow H_c}$ depends on both the 
fragmentation parameter $z$ and the factorization scale $\mu$.
The function (\ref{last}) should be regarded as a model for the gluon FF at the scale $\mu_0$ of 
order $2 m_c$.
For values of $\mu$ much larger than $\mu_0$, the obtained FF should be evolved from the
scale $\mu_0=2 m_c$ to the desired scale $\mu$ using the DGLAP equation (\ref{dglap}).\\
In Fig.~\ref{plot}, the FF of $g\rightarrow H_c$ at the starting scale $\mu_0=2m_c=3$~GeV is shown. 
The behavior of $D_g^{H_c}$ is shown for different values of the transverse momentum of the gluon.\\
Note that one of the uncertainties in determination of FFs is due to 
the freedom in the choice of scaling variable $z$, i.e. the covariant and non-covariant definitions 
presented in Eqs.(3)-(7). As a comparison, our result shown in Fig. \ref{plot} 
is in reliable consistency  with the result presented in Fig. 3 of Ref.\cite{Qi:2007sf} when  the third 
definition of fragmentation parameter  (Eq. (5)) is applied, i.e. $z=\sqrt{M_H^2+(k_L^H)^2}/E^g=E^H/E^g$.
Since there are no phenomenological results for the $g\rightarrow H_c$ FF, then
at the moment it is not possible for us to compare our result phenomenologically, but
the behavior of our obtained FF in comparison with the $g\rightarrow B$ and $g\rightarrow D^+$ FFs, shown in 
Figs. \ref{bmeson} and \ref{dplusmeson} assures one to rely on our result. 

\section{Conclusion}
\label{sec:four}

The dominant production mechanism for heavy quarkonium at high
transverse momentum is fragmentation; the production of a high energy parton followed by its splitting into 
the heavy quark-antiquark bound states. 
We showed that the fragmentation function which  describes this process
can be calculated using perturbative QCD. 
In this work we gave out the initial fragmentation functions of gluon  
to split into S-wave charmonium states to leading order in $\alpha_s$.
We used a different approach,  Suzuki's approach, in getting them from
the normal analytic calculation in the literatures \cite{Qi:2007sf,Braaten:1993rw}
and found good agreement with the result in \cite{Qi:2007sf} when use the normal 
definition of the fragmentation parameter, i.e. $z=E^H/E^g$.
Finally, although the  fragmentation function obtained in this work (\ref{last}) is schematically for charmonium, in fact 
it can be directly applied to the S-wave bottomonium sate $\Upsilon$ except 
that $m_c$ is replaced by $m_b$ and the decay constant $f_M$ is the appropriate constant for the bottomonium mesons.

\end{document}